\documentclass[useAMS,usenatbib]{mn2e}

\usepackage{graphicx,amssymb,bm}
\usepackage{subfig,xfrac}
\usepackage{float, Abbrev}
\usepackage[fleqn]{amsmath}  
\usepackage{color}

\usepackage{multirow}

\newcommand{\be}{\begin{equation}}
\newcommand{\ee}{\end{equation}}
\newcommand{\ba}{\begin{eqnarray}}
\newcommand{\ea}{\end{eqnarray}}

\def\simlt{\lower.5ex\hbox{$\; \buildrel < \over \sim \;$}}

\newcommand{\fig}{\begin{figure} \begin{center}}
\newcommand{\efig}{\end{center}\end{figure} }
\newcommand{\figs}{\begin{figure*}\begin{minipage}{180mm} \begin{center}}
\newcommand{\efigs}{\end{center}\end{minipage}\end{figure*} }
\def\simgt{\lower.5ex\hbox{$\; \buildrel > \over \sim \;$}}

\usepackage{graphicx} 

\title[Where is the BCG?]{A detection of wobbling Brightest Cluster Galaxies within massive galaxy clusters}
\author[D. Harvey et al]
{David Harvey$^{1}$\thanks{e-mail: {\tt david.harvey@epfl.ch}}, F. Courbin$^{1}$, J. P. Kneib$^{1,2}$, Ian G. McCarthy$^{3}$\\
$^{1}$Insitute of Physics, Laboratoire d'Astrophysique, EPFL, Observatoire de Sauverny, 1290 Versoix, Switzerland \\
$^{2}$Aix Marseille Université, CNRS, LAM (Laboratoire d'Astrophysique de Marseille) UMR 7326, 13388, Marseille, France\\
$^{3}$Astrophysics Research Institute, Liverpool John Moores University, 146 Brownlow Hill, Liverpool L3 5RF}
\begin{document}

\date{Accepted ---. Received ---; in original form \today.}

\pagerange{\pageref{firstpage}--\pageref{lastpage}} \pubyear{2017}

\maketitle

\label{firstpage}

\begin{abstract}
% The collisionless cold dark matter paradigm predicts that the Brightest Cluster Galaxy  (BCG) should lie coincident with the centre of a relaxed galaxy cluster.
A striking signal of dark matter beyond the standard model is the existence of cores in the centre of galaxy clusters.
Recent simulations predict that a Brightest Cluster Galaxy (BCG) inside a cored galaxy cluster will exhibit residual wobbling due to previous major mergers, long after the relaxation of the overall cluster.
This phenomenon is absent with standard cold dark matter where a cuspy density profile keeps a BCG tightly bound at the centre.
We test this hypothesis using cosmological simulations and deep observations of 10 galaxy clusters acting as strong gravitational lenses.
% and ask if there exists any evidence for excessive variance in the offset between the BCG and the large scale cluster halo that cannot be accounted for with purely positional uncertainty. 
%To do this we first generate a suite of simulations that are constructed to closely reflect our sample.
%With the halos directly coincident with their BCGs, we empirically estimate the uncertainty the position of the cluster halo finding that on average the each cluster has a radial error in the position of $\sigma_r\simeq0.7\arcsec$, compared to the uncertainty returned by {\sc Lenstool} of $\sigma_{\rm r}\simeq0.05\arcsec$. 
%Following this test we generate four more suites of simulations, each with varying wobble amplitude $A_{\rm w}$. 
%We then compare the distribution of best fitting positions from each suite with numerically calculated posteriors and calculate the Kolmogorov-Smirnov two-sample probability.
%We find that the predicted $A_{\rm w}$ for each simulation is consistent with the input value at $1\sigma$.
Modelling the BCG wobble as a simple harmonic oscillator, we measure the wobble amplitude, $A_{\rm w}$, in the BAHAMAS suite of cosmological hydrodynamical simulations, finding an upper limit for the CDM paradigm of $A_{\rm w} < 2$kpc at the 95\% confidence limit.
We carry out the same test on the data finding a non-zero amplitude of $A_{\rm w}=11.82^{+7.3}_{-3.0}$kpc, with the observations dis-favouring $A_{\rm w}=0$ at the $3\sigma$ confidence level.
This detection of BCG wobbling is evidence for a dark matter core at the heart of galaxy clusters. 
It also shows that strong lensing models of clusters cannot assume that the BCG is exactly coincident with the large scale halo.
While our small sample of galaxy clusters already indicates a non-zero $A_{\rm w}$, with larger surveys, e.g. Euclid, we will be able to not only to confirm the effect but also to use it to determine whether or not the wobbling finds its origin in new fundamental physics or astrophysical process.
 
\end{abstract}

\begin{keywords}
cosmology: dark matter --- galaxies: clusters --- gravitational lensing
\end{keywords}

\section{Introduction}
Cosmological simulations of the Universe predict that structure should form a web like texture \citep[e.g.][]{EvolutionLSS,illustris,millennium,eagle}.
Lying at the nodes of the cosmic web are the largest known structures in the Universe.
Galaxy clusters consist of thousands of galaxies, all embedded within a plasma of hot X-ray gas and a halo of dark matter \citep[for a review see][]{galaxy_cluster_review}.

It is not uncommon for the mass of a galaxy cluster to exceed $M=10^{14}M_\odot$ or even $M=10^{15}M_\odot$ \citep[e.g.][]{A2744,A2744_HFF,MACSJ1149_HFF,MACSJ0717_HFF,A1689,A2163,MACSJ0416,harvey_0416}. 
In these environments the curvature of space-time becomes increasingly warped, causing objects that lie behind these clusters to be distorted.
Should objects such as galaxies serendipitously find themselves directly in the line of sight of the observer and the cluster, their image can be split in to multiple images of the same distant, `source' galaxy \citep[for review, please see e.g.][]{gravitational_lensing}.
Using these multiple images of the same source, we can reconstruct the distribution of foreground matter \cite[e.g][]{Locuss_Richard,CLASH_zitrin,Merten_clash,HFF_recons}.

Despite the success of the non-relativistic (cold), collisionless dark matter paradigm at predicting  the large scale distribution of galaxies within the Universe \citep[e.g.][]{EvolutionLSS,BOSS_clustering,2dgf,vipers}, small scale discrepancies between observations and data means that tensions still exist. Observations of the local group have highlighted issues regarding the under-abundance of dwarf galaxies, plus the apparent cored density profiles of these galaxies, something not predicted by CDM \citep{toobigtofail,corecusp,lostsatellite,lostsatellite1,corecusp,corecusp1}.
Although solutions to reconcile these discrepancies could lie within complex baryonic physics \citep[e.g.][]{baryonsolution,AGNfeedback1}, it is possible to alleviate this tension by invoking more exotic forms of dark matter that have extra degrees of freedom, for example warm dark matter \citep{WDM,WDM1,WDM2} or self-interacting dark matter \citep[e.g.][hereafter SIDM]{HaloSIDM,SIDMSim,SIDMSimA,SIDMSimB}. 

The ability to reconstruct the total mass of a cluster to within 1\% has now become common place with a multitude of different algorithms allowing both parametric and free form fits to the data \citep[e.g][]{merten_SL,lenstool,strongweakunited,strongweakunited1,strongweakunited2}.
This, in conjunction with the availability of high resolution, deep imaging from space \citep{CLASH,HFF}, means that we now have the ability to accurately reconstruct the distribution of dark matter that lies within galaxy clusters along with its baryon counterparts and hence have now become important test-beds for the cold dark matter paradigm. 

Initial studies of CDM within clusters have provided unequivocal proof for its existence when it separates from the gas component during mergers \citep{separation,bulletcluster,bulletclusterA,bulletclusterB,minibullet,musket}. Moreover they have become direct tests for the collisionless assumption associated with CDM \citep{SIDMTest,SubhalosSIDMA,ObserveSIDM,Harvey14,Harvey15,A3827_massey,cannibal,substructure_a2744}. Since these studies there have been further investigations looking at different ways in which we can study the cold dark matter paradigm using galaxy clusters, for example looking for misalignments between galaxies and their dark matter halo in the Hubble Frontier Fields \citep{Harvey16}, trailing dark matter after a collision \citep{Harvey_trails,dmtrails_2} and the wobbling of the Brightest Cluster Galaxy (BCG) within the cluster centre \citep{darkgiants}.

Where the centre of a group or cluster of galaxies has remained an unanswered question in the literature. It is usual that in the centre of these galaxy clusters lies a giant galaxy, known as the Brightest Cluster Galaxy (BCG). \cite{galgroup} did an exhaustive study of what the best estimate of the centre of a group of galaxies was, over a mass range of $M=(10^{13}-10^{14})$M$_\odot$. Using the weak lensing profile as a proxy for `goodness of fit' test for the centroid, they found that the most massive galaxy near an X-ray peak was the best identifier for the centre of a galaxy group.  However this study still found some evidence for an offset between the weak lensing centre and the most massive galaxy. 

In clusters of galaxies where the mass range is an order of magnitude higher, the question becomes very different. When reconstructing the mass profile of galaxy clusters with parametric lens modelling tools, it it is often assumed that the centre of the large scale dark matter to be kept fixed on the BCG. This is from the fact that in a CDM paradigm the BCG well traces the bottom of a steep gravitational potential, with no known process that could offset a galaxy from its halo \citep{LCDM_offsets}. However, \cite{darkgiants} found that in the event that dark matter self-interacts, then a relaxed cluster could exhibit some residual wobbling  due to the existence of a remnant core in the density profile \citep{darkgiants}. Any detection of a wobble would infer the existence of a core, and hence some potential new physics beyond the standard model. 

The debate of whether a core exists in galaxy clusters continues. \cite{densityProf2,densityProf3} carried out a detailed study of seven massive relaxed galaxy clusters to see if any core existed by direct measurement of the total mass distribution. Using a combination of stellar dynamics, strong lensing and weak lensing there directly measured a mean core of $\langle\log r_{\rm core}\rangle=1.14\pm 0.13$~kpc. They found that although the density profile of the dark matter seemed to lower in the core than what was expected from a typical NFW profile \citep{NFW}, this was correlated with the distribution of stars. Although suggestive of their existence, it still remains an unanswered question of whether clusters, like dwarf galaxies, exhibit central cores.

In this paper we study a sample of 10 galaxy clusters and ask the question `does the bright cluster galaxy lie at the centre of the large scale dark matter potential or does evidence exist for some finite wobbling?' We will carry out the same measurement on galaxy clusters from the BAHAMAS simulations \citep{BAHAMAS} and compare the predictions of cold dark matter to observations.

In section \ref{sec:data} we will outline the suite of cosmological simulations and the observations of the galaxy clusters used in this study. 
In section \ref{sec:method} we outline our mass reconstruction method using strong gravitational lensing, and how we estimate the wobbling of the BCG in section \ref{sec:osc_amp}.
In section \ref{sec:results} we present our results from the simulations and observations and in
section \ref{sec:conclusions} we conclude and discuss future observations.
Throughout the paper we use a cosmology of $\Omega_{\rm M} =0.3$, $\Omega_{\rm \Lambda} =0.7$ and $h=0.7$ to convert from angular separation to angular diameter distances.

\begin{table}
\centering
\begin{tabular}{|c|c|c|}
\hline
Parameter & Mean Lenstool Error &  Empirical Error \\\hline
\hline
$\sigma_{\rm x}$ & $0.05\arcsec$ & $0.3\arcsec$   \\
$\sigma_{\rm y}$ & $0.04\arcsec$ & $0.6\arcsec$    \\ 
$\sigma_{\rm radial}$ &  $0.05\arcsec$ &  $0.7\arcsec$   \\ 
$\sigma_{\rm M_{\rm 200}}/M_{\rm 200}$ & 0.1 dex & 0.2 dex \\ 
$\sigma_{\rm c_{\rm 200}}/c_{\rm 200}$ & 0.07 dex  & 0.55 dex   \\ \hline
\end{tabular}
\caption{\label{tab:errors} We calibrate our errors empirically and do not use those derived from the width of the posterior in the MCMC. We run a simulation with the main halo and BCG coincident, with a multiple image random error of $\sigma_{\rm im}=0.5\arcsec$. We then calculate the scatter around the true value and use these as the uncertainty on the position throughout the analysis.}
\end{table}
\begin{table*}
\centering
\begin{tabular}{|c|c|c|c|c|c|c|c|c|c|c|c|}
\hline
Cluster & RA & DEC & $z$ & $\Gamma$ & N$_{\rm I}$ & N$_{\rm P}$ & RMS & $\delta_{x}$ ($\arcsec$) & $\delta_{y}$($\arcsec$) & $M_{200} (\times10^{14}M_\odot)$ & $c_{200}$\\
\hline
A1063 & 342.18324 & -44.53087 & 0.35 &0.24 & 41 & 17 & 0.88 & $0.05\pm0.12$ & $-0.01\pm0.10$ & $16.2\pm0.3$ & $4.7\pm0.1$ \\ %\hline
A383 & 42.01409 & -3.52938 & 0.19 & 0.53 & 26 & 11 & 0.75 & $0.13\pm0.18$ & $0.68\pm0.15$ & $16.6\pm5.1$ & $4.0\pm1.4$ \\ %\hline
A2261 & 260.61341 & 32.13266 & 0.22 & 0.36 & 32 & 20 & 0.68 & $0.67\pm0.04$ & $-0.56\pm0.05$ & $6.9\pm0.6$ & $9.0\pm0.5$ \\ %\hline
A1703 & 198.77197 & 51.81749 & 0.28 & 0.20 & 42 & 22 & 1.03 & $-0.55\pm0.09$ & $1.07\pm0.16$ & $13.5\pm0.9$ & $4.6\pm0.2$ \\ %\hline
A1835 & 210.25865 & 2.87847 & 0.25 & 0.51 & 18 & 14 & 1.20 & $4.25\pm0.06$ & $-0.64\pm0.12$ & $28.7\pm1.7$ & $3.7\pm0.2$ \\ %\hline
A1413 & 178.82449 & 23.40445 & 0.14  & 0.33 & 11 & 9 & 0.75 & $0.07\pm0.06$ & $-0.71\pm0.13$ & $6.7\pm0.9$ & $7.2\pm0.5$ \\ %\hline
MACS0744 & 116.21999 & 39.45740 & 0.69 & 0.38 & 20 & 19 & 1.52 & $-0.62\pm0.21$ & $-0.42\pm0.61$ & $9.9\pm1.4$ & $4.7\pm0.8$ \\ %\hline
MACS1206 & 181.55060 & -8.80093 & 0.44 & 0.28 & 35 & 13 & 1.62 & $-1.36\pm0.07$ & $-0.36\pm0.03$ & $15.0\pm0.2$ & $4.8\pm0.1$ \\ %\hline
MACS1720 & 260.06980 & 35.60731 & 0.39 & 0.44 & 17 & 13 & 1.61 & $0.61\pm0.04$ & $-1.64\pm0.05$ & $9.8\pm0.7$ & $5.2\pm0.3$ \\ %\hline
MACS1931 & 292.95683 & -26.57584 & 0.35 &0.55 & 23 & 10 & 0.91 & $0.45\pm0.03$ & $1.91\pm0.09$ & $9.7\pm0.3$ & $5.0\pm0.1$ \\ \hline
\end{tabular}
\caption{The survey sample of the ten dynamically relaxed galaxy clusters in which we aim to measure the offset between the BCG and large scale main halo. {\it Col 2:} Right Ascension, {\it Col 3:} Declination, {\it Col 4:} Cluster redshift, {\it Col 5:}  Dynamical state (eq. \eqref{eqn:dynamical}), {\it Col 6:}  Number of multiple images, {\it Col 7:} Number of parameters in the fit {\it Col 8:} Root mean square error of the mode fit, {\it Col 9 \& 10:} the offset between the BCG and the large scale halo in arc-seconds, {\it Col 11:} Mass of the cluster halo and {\it Col 12:} the concentration parameter. All uncertainties are purely statistical Gaussian errors as reported by {\sc Lenstool}.
\label{tab:data}}
\end{table*}

\section{Data}\label{sec:data}

Here we describe our two primary samples of data. The sample of clusters from our N-body simulations, and the observations using the Hubble Space Telescope (HST).

\subsection{N-body simulations of galaxy clusters}

We use a number of simulations from the BAHAMAS suite of cosmological hydrodynamical simulations \citep{BAHAMAS} to predict the offset distribution for the standard $\Lambda$CDM scenario (i.e., non-interacting DM).  BAHAMAS consists of large 400 Mpc/$h$ on a side periodic box simulations with 1024$^3$ baryon and CDM particles and a force softening of 4 kpc/$h$, run in a number of different background cosmologies.  Here we use a set of four independent realisations of a WMAP9 cosmology, along with a higher-resolution 100 Mpc/$h$ box with 512$^3$ particles and a softening of 2 kpc/$h$ (also WMAP9), to test for a resolution dependence in the predicted offset distribution.

BAHAMAS was run using a modified version of the {\sc Gadget 3} code (last described in \citealt{gadget2}).  The simulations include sub-grid treatments for metal-dependent radiative cooling, star formation, stellar evolution and chemodynamics, and stellar and active galactic nuclei (AGN) feedback, developed as part of the OverWhelmingly Large Simulations project (see \citealt{OWLS} and references therein).  \citet{BAHAMAS} calibrated the stellar and AGN feedback to reproduce the local galaxy stellar mass function and the amplitude of the gas mass$-$halo mass relation of galaxy groups and clusters, as inferred by spatially-resolved X-ray observations.  As demonstrated in \citet{BAHAMAS}, the simulations reproduce not only the overall gas and stellar content of groups and clusters, but also the detailed radial distributions of these components and the observed split in stellar mass between satellites and centrals.  

The large volume of the BAHAMAS runs allows us to extract a large statistical sample of galaxy clusters for comparison with our observed sample of massive clusters.  From the four 400 Mpc/$h$ boxes, we select all haloes from the $z=0.25$ snapshot with $M_{200} > 3\times10^{14}$ M$_\odot$, yielding a combined sample of $\approx600$ systems.  For each cluster, we produce three maps: i) a total surface mass density map (stars+gas+CDM); ii) a stellar surface mass density map; and iii) an X-ray surface brightness map.  All three maps are centred on the potential minimum and span a field of view of 2 Mpc (physical) with a pixel resolution of 1 kpc.  The maps are created by summing mass (or X-ray emission) along the line of sight, using a column length of 10 Mpc (i.e., particles with $|z-z_{\rm min}| < 5$ Mpc are selected, where $z_{\rm min}$ is the z-coordinate of the potential minimum).  Particles are mapped to the grid using
a smoothed particle hydrodynamics interpolation scheme with 24 neighbours.

\subsection{Selecting a uniform sample of clusters from simulations and observations}
When selecting our clusters from the simulations we want to ensure that primarily that the clusters are relaxed, plus they have the same selection function as the observations. To this extent we ensure X-ray gas has a uni-modal distribution and have a dynamical state parameter $\Gamma\ge0.2$, where $\Gamma$ is defined as the ratio of the X-ray flux, $S$, within 100 kpc and 400 kpc i.e.
\be
\Gamma = \frac{S(<100{\rm~kpc})}{S(<400{\rm~kpc})}.
\label{eqn:dynamical}
\ee
This definition was previously shown to be the most robust measure for the dynamical state of a galaxy clusters using high-resolution hydrodynamical zoom simulations of clusters \citep{dynamical_state_xray}.

Applying this relaxation criterion results in a total of 190 clusters in our final sample.  However, we have checked that our results are not sensitive to the specific relaxation criterion adopted.

We extract the the position of the BCG for the stellar distribution by running SExtractor \citep{sextractor} on the stellar surface mass density maps, to try to mimic the method used for the real data.  For the centre of the dark matter, we use the deepest part of the potential (of the total matter).
We then measure the offset between the measured BCG position as derived from SExtrator and the centre of the potential.

For the observations we use the X-ray emission maps from the Chandra X-ray telescope (for information regarding the reduction and extraction please see \cite{Harvey15}).

\subsection{HST observations of relaxed galaxy clusters}
We analyse clusters from the Local Cluster Substructure Survey \citep[][LoCuSS]{Locuss_Richard} and the Cluster Lensing And Supernova survey with Hubble (CLASH) 
we select only those clusters that have greater than or equal to 10 {\it confirmed} multiple images at differing redshifts. 
This equates to a sample of 10 galaxy clusters.

The selected multiple images are those presented in \cite{CLASH_zitrin}, \cite{Locuss_Richard} and \cite{a1703}.
We also use the cluster member catalogues from  \cite{CLASH_zitrin}, \cite{Locuss_Richard} and \cite{a1703}, who used the red sequence to identify each member. 
We derive the luminosity of the cluster members in the CLASH clusters using the publicly available photometric catalogues \citep{CLASH_photoz}.
We also use the the photometric redshifts from the same catalogues for any multiple images that do not have spectroscopic redshifts and used their associated error as a Gaussian prior \citep{CLASH_photoz}, hence keeping these source redshifts a free parameter within the range allowed by the prior.

\section{Method}\label{sec:method}

To measure whether the BCG oscillates around the centre of a relaxed galaxy cluster we use strong gravitational lensing to reconstruct the total mass distribution.
In order to carry out a systematic, uniform measurement over the sample of galaxy clusters, we attempt to use the same methodology for each cluster. 

\subsection{Strong lensing modelling}
We model the distribution of matter in each galaxy cluster using the open source software {\sc Lenstool} \citep{lenstool}, which is a strong lensing Bayesian algorithm that uses analytical fits.
It then samples this parameter space using a Monte Carlo Markov Chain (MCMC) with Metropolis Hastings sampling returning the estimated posterior for each parameter.
We fit the main large scale, cluster halo with a NFW profile \citep{NFW}, with mass, concentration, ellipticity, position angle and x-y position as free parameters (making a total of six for the main halo). We also apply a flat prior on the position of the halo with a width of 10 arc-seconds in the x and y position. We address the dependency of this choice later in the study.

We then assume that each galaxy member lies on the fundamental plane \citep{Locuss_Richard} and that each galaxy scale halo can be described by a pseudo isothermal elliptical mass distribution (PIEMD). The PIEMD follows the radial density profile
\be
\rho_{\rm PIEMD} = \frac{\rho_0}{(1+r^2/r_{\rm core}^2)(1+r^2/r_{\rm cut}^2)},
\ee 
where $r_{\rm core}$ and $r_{\rm cut}$ are the core and cut radius respectively, where we assume 
\be
r_{\rm core}=r_{\rm core}^\star\left(\frac{L}{L_\star}\right)^{1/2},
\ee
\be
r_{\rm cut}=r_{\rm cut}^\star\left(\frac{L}{L_\star}\right)^{1/2},
\ee
\be
\rho_0 = \frac{\sigma_0^2}{2\pi {\rm G}}\frac{(r_{\rm core} + r_{\rm cut})}{(r_{\rm core}^2r_{\rm cut})},
\ee
and the 3D velocity dispersion, $\sigma$, of the PIEMD is
\be
\sigma=\sigma^\star\left(\frac{L}{L_\star}\right)^{1/4}.
\ee
Following \cite{Locuss_Richard}, we assume that the normalisation of the galaxy masses scale with their luminosity, relative to a $L^\star$ galaxy, $r_{\rm core}^\star=0.15$ kpc, and we free up the cut radius with a tight Gaussian prior $r_{\rm cut}^\star=45\pm1$ kpc and the velocity dispersion as $\sigma^\star=158\pm27$ km/s.
In some very specific cases where there happens to be a multiple image wrapped around an individual galaxy member, we free up a galaxy velocity dispersion and cut radius as well. 
\fig
\includegraphics[width=0.45\textwidth]{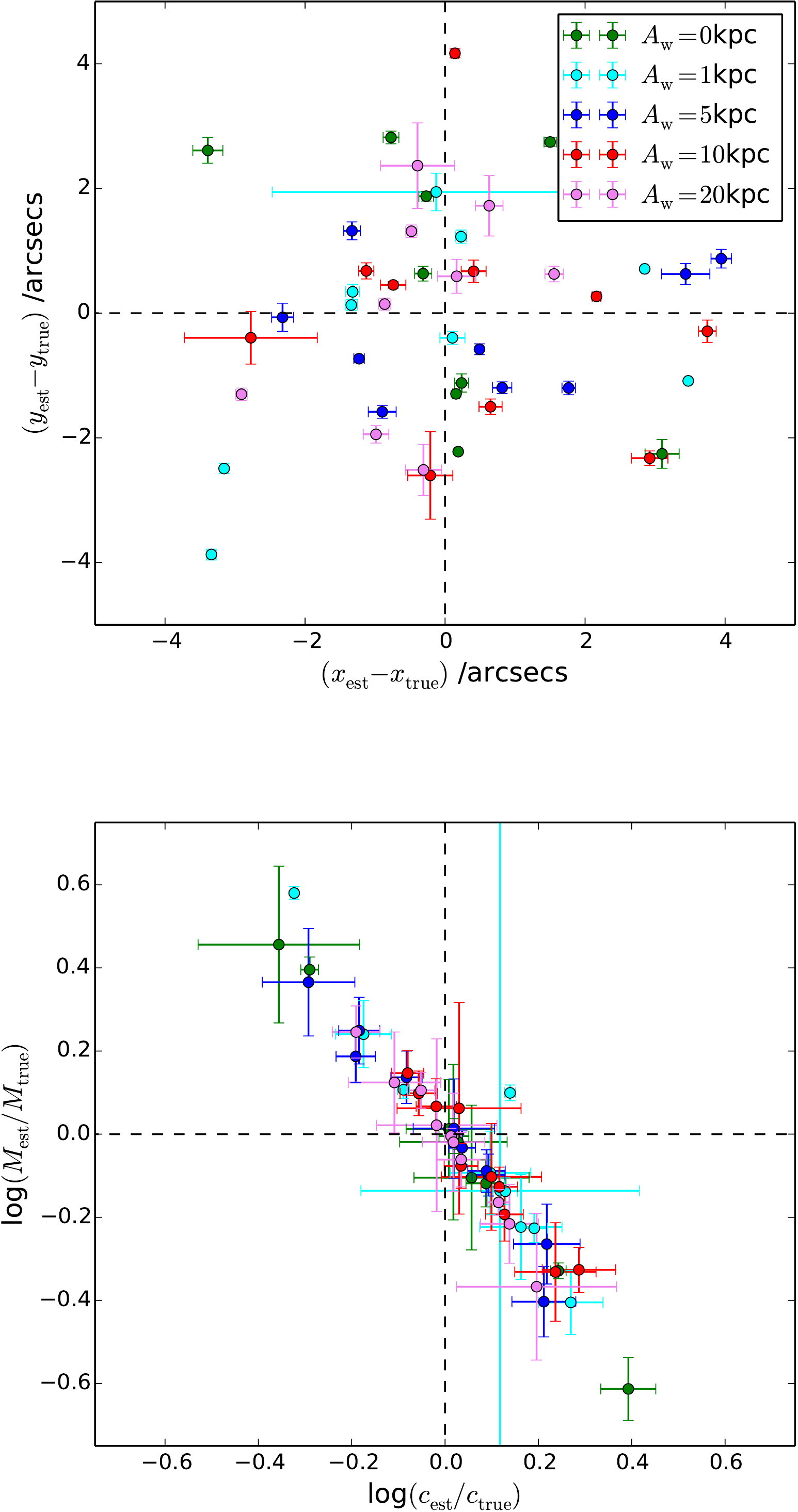}
\caption{\label{fig:sims} We run five suites of simulations each with a different wobble amplitudes, $A_{\rm w}$ to calibrate our positional uncertainties. The top panel shows the offset in position between the estimated and true position of large scale cluster halo (in arc-seconds) and the bottom shows the logarithm of the ratio between the estimate and true value of the concentration and mass of the large scale NFW halo. In both cases the error bars are the marginalised ones derived from the {\sc Lenstool} posterior. It can be seen that the statistical scatter in the positional offsets is much larger than the estimated uncertainties.}
\efig

\fig
\includegraphics[width=0.45\textwidth]{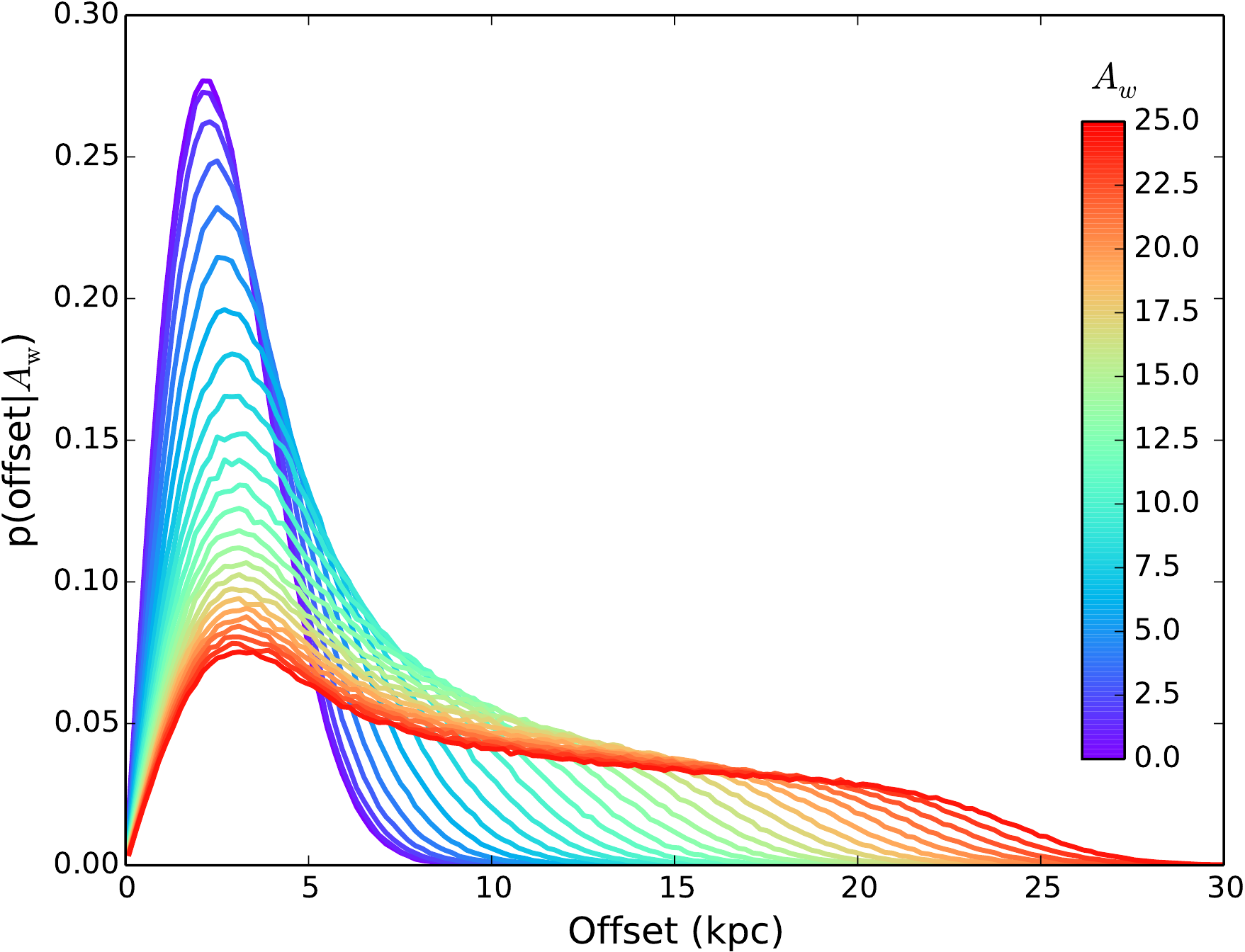}
\caption{We calculate $p({\rm offset}| A_{\rm w})$, the expected posterior for the radial offset between BCG and cluster halo, given some wobble amplitude, $A_{\rm w}$ kpc. To do this we generate them numerically folding in the true redshift distributions of the clusters (see text). The kink in the tail of the posterior is due to the hard cut off in the simple harmonic oscillation convolved with statistical errors in position.
\label{fig:posteriors}}
\efig	

\subsection{Mass of the Bright Cluster Galaxy (BCG)}

The inherent degeneracy of the large scale halo and the BCG in the very centre of the cluster means that we do not have constraints on its mass. 
We therefore choose to model the BCG in the cluster as part of the galaxy members, assuming they lie along the fundamental plane. This was found to be accurate, and in fact, the BCG had less scatter than that of normal early type galaxy \citep{BCG_fundamentalplane}.
%This may result in an underestimate of the mass of the BCG and hence over estimate of the mass of the large scale halo, however should this be the case, it will bias the main halo towards the BCG and cause an an under-estimate of the wobble of the BCG. Therefore any non-zero finding will have a under-estimate of its significance.
%In an initial test we test this carry out the reconstructions attempting to constrain the velocity dispersion and cut radius of the BCG. Although we find no 
Future work, with potential integral field spectroscopic observations of the BCG will allow us to break this degeneracy, allowing us to exploit larger sample sizes.
% \com{Fred: you could also say that you don't care too much about the details of the mass profile as long as the centroid is not affected by the degeneracies. On the other hand it would be nicer to be more quantitative here, with an estimate of the degeneracy and of the maximum possible bias on the centroid.}

\subsection{Mass reconstruction}
Following the methodology above we present the results from our sample of 10 galaxy clusters. Table~\ref{tab:data} shows an overview of the data and the fits. 
For each galaxy cluster we give 
the right ascension (RA) and declination (DEC) of the BCG, 
the dynamical state of the cluster according to equation~\eqref{eqn:dynamical},
the number of multiple images used in the analysis and the number of parameters in the fit, including free redshifts,
the root mean square (RMS) of the true image position to the predicted image position
and then best fit position (with respect to the BCG), the mass and concentration of the cluster scale halo. 
In each case the error bar is the statistical marginalised error bar returned by {\sc Lenstool}, i.e. the $1\sigma$ width of the posterior.
We find that our models have very similar RMS to previously found for these clusters (for example \cite{CLASH_zitrin}), lending confidence to our models.

\subsection{Empirical measurement of positional errors with image simulations}\label{sec:errors}
\figs
\includegraphics[width=0.8\textwidth]{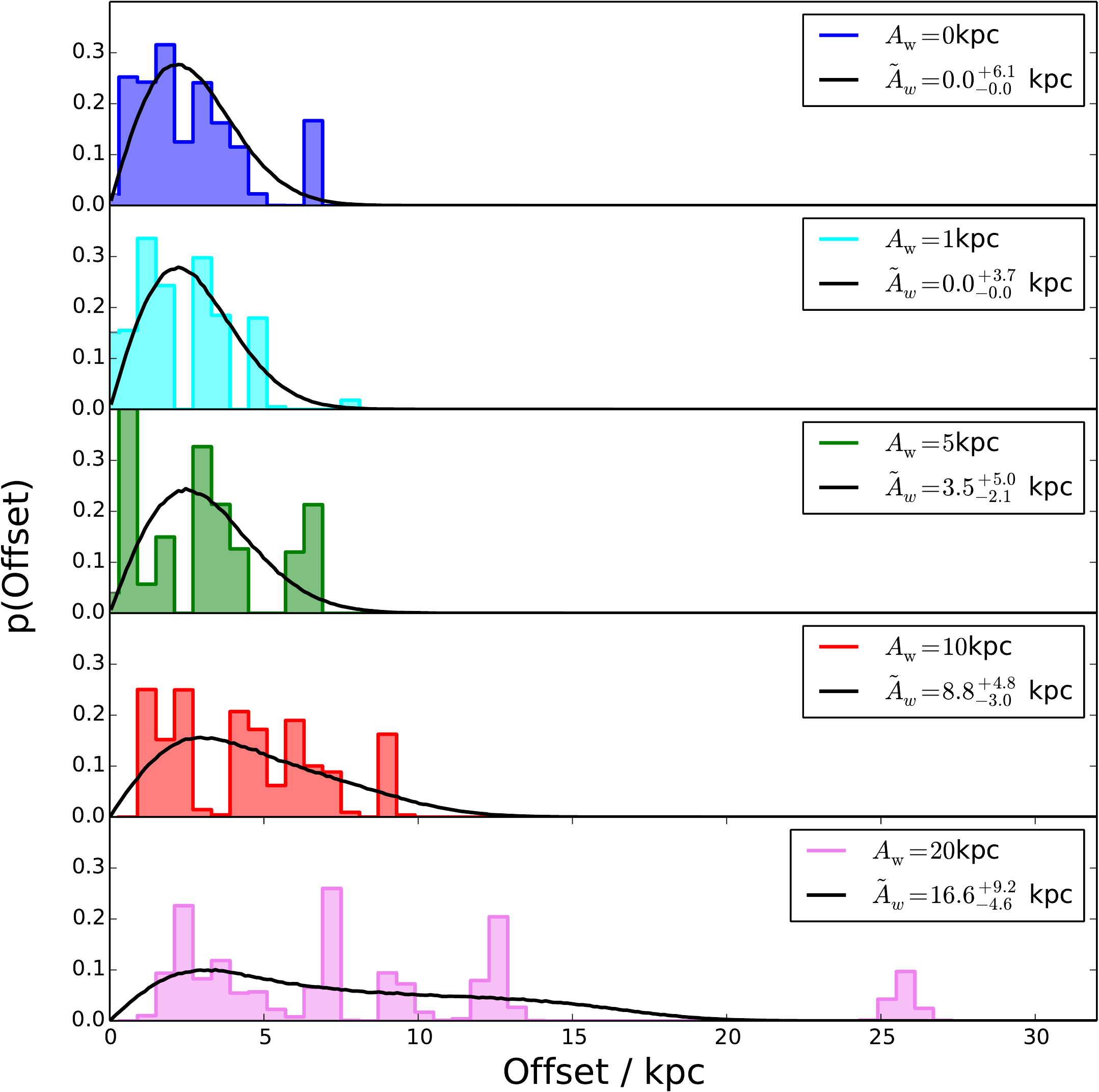}
\caption{\label{fig:compare_posteriors} 
Each panel shows the stacked radial posteriors for a different image simulation.
Each simulation has 10 galaxy clusters with a different input BCG wobbling amplitude centred around the main cluster halo (coloured histogram), with the key showing the input wobble amplitude $A_{\rm w}$.
Overlaid is the best fitting posterior from Figure \ref{fig:posteriors}. The corresponding estimated wobble amplitude,  ${\tilde A}_{\rm w}$ with its associated uncertainty is shown in the key. In each case the estimated wobble amplitude is consistent with the input, true wobble amplitude}
\efigs
The errors within {\sc Lenstool} are  derived from the width of the posterior sampled during {\sc Lenstool's} MCMC. 
%However, these are often purely statistical and they depend somewhat on the user input expected root mean square error in the photometry of the multiple positions. 
These therefore do not account for the systematic errors that could reside within the reconstruction. In order to understand better the total error budget for each cluster we empirically measure the errors  by running a suite of image simulations.
Each image simulation is based on the actual data. It uses the expected source positions from each actual cluster, then using a known cluster model (centred exactly on the BCG) we create a set of multiple images such that it mimics the observations.
Therefore simulating lensed images that are close to the observed positions. For each cluster we simulate a circularly symmetric NFW halo, with a mass of $M_{\rm NFW} = 10^{15}{\rm M}_\odot$ (given that we find the mean mass of the actual data to be $\langle M_{\rm 200}\rangle=1.3\times10^{15}{\rm M}_\odot$ and hence will slightly over-estimate our error bars). The concentration follows that of \cite{massConc3} for a cluster mass of $M = 10^{15}{\rm M}_\odot$ (and redshift of each cluster), and using the same cluster member catalogue with the same $L_\star$ value, we adopt the central values of the prior as outlined in section \ref{sec:method}. We then randomly move the simulated multiple images according to the statistical error input in to the reconstruction (using the expected RMS $= 0.5\arcsec$ from the actual reconstructions), and analyse the cluster in exactly the same way we do the data. Since we know the simulated large scale halos are exactly centred on the BCG, this Monte Carlo approach of estimating the error bars allows us to estimate the error in the actual data, better reflecting the total systematic plus statistical error. In other words, this strategy allows us to estimate the total measurement errors inherent to the lensing reconstruction in the absence of any wobbling.

%Hence this will allow us to quantify any excess variance in the offsets observed that is not consistent with just observational noise. 
We then repeat this experiment with image simulations four more times.
Each time we introduce a random offset drawn from a known calculated posterior, in to the BCG, mimicking a wobble.
We simulate five different wobble amplitudes, $A_{\rm w}=[0, 1, 5, 10, 20]$~kpc (see section \ref{sec:osc_amp}).
Given we know the true input position we use all five simulations to estimate the total error in the mass reconstruction.
Figure \ref{fig:sims} shows the results of our five sets of image simulations, each consisting of 10 clusters. The top panel shows the positional estimates of the large scale halo in each image simulation with respect to the simulated true position. The error bars are the Gaussian uncertainties derived directly from the width of the posterior in {\sc Lenstool}. In some cases it can be seen that the uncertainties in the position are too small compared to the true position, which is why we carried out a Monte Carlo estimate of the total error bars. The bottom panel shows the logarithm of the ratio of the estimated and true input mass and concentration. The degeneracy between these two parameters is clear, suggesting that point estimates of mass and concentration are biased and replaced by the full 2D posterior distributions for the parameters.

Despite this, we find that {\sc lenstool} predicts an average error $\sigma_x =0.05\arcsec$,$\sigma_y=0.04\arcsec$ and a propagated error in the radial offset of $\sigma_r =0.05\arcsec$,. 
However the scatter of the estimated positions gives an error of $\sigma_x=0.3\arcsec$, $\sigma_y=0.6\arcsec$  and $\sigma_r=0.7\arcsec$. This equates to a factor of 10 difference between what {\sc Lenstool} predicts the error is, and what the true error on the position is. We therefore adopt Monte Carlo error estimates as these are empirical and based on the data and hence more reliable.

Following the calibration of the uncertainty in the position, we use the reconstructed clusters from the simulated multiple images to see if we can recover the input wobble amplitudes introduced into these simulations. 

\subsection{Estimating the wobble amplitude}\label{sec:osc_amp}

To calculate the input wobble we assume that if the BCGs are oscillating about the cluster core, they do so in simple harmonic motion. 
This is a simplification but enables us to predict the distribution of 2D offsets when their 3D motion is viewed in projection.
In order to estimate the amplitude of the BCG wobble, $A_{\rm w}$, we numerically generate posterior probability distributions for multiple $A_{\rm w}$ and then compare these posteriors to the data to find the best fitting posterior and its associated wobble amplitude. 
\cite{darkgiants} found that over the simulated period post merger ($\sim6$Gyr), the wobble amplitude was the same size as the core and behaved like a minimally decaying simple harmonic oscillator. We assume in this study that there is no damping, however this may not be true and simulations will need to test this. However, this is beyond the scope of this paper.
To generate posteriors as a function of wobble amplitude, we model the wobble as a simple harmonic oscillator, where the radial distance the BCG is from the centre of the potential follows the normal solution,
\be
r(t) = A_{\rm w}\cos(\omega t),
\ee
where  $w$ is inversely proportional to the period of the oscillation and $t$ is the time. Since we expect the oscillation to be periodic we generate multiple realisations of the radial distance (in 3-dimensions which is then projected) to create an expected probability distribution function of radial offsets on the plane of the sky. We then add to this statistical error in the position of the BCG, which we calibrate empirically.

Using the extracted error estimates, we fold these into the estimate of the expected distribution of radial offsets for a given wobble amplitude, $A_{\rm w}$, and random error, $\sigma_{r}$.
The posteriors calculated for various oscillating amplitudes, $A_{\rm w}$ are shown in Figure \ref{fig:posteriors}.
We then compare the distribution of best fitting positions from simulations to these posteriors and generate a probability that the distribution of positions were pulled from that posterior using the Kolmogorov-Smirnov two sample test.

%Since during the fit of the NFW halo, the MCMC samples in x-y space, and this is then converted in to polar coordinates, the resulting distribution will be a Rayleigh-like. However, given that the method is susceptible to additional sources of noise from the randomly sampled redshift of the cluster (and hence the error in proper distance will be different for each, even if each cluster has an error of 0.5 arc-seconds) we generate the posteriors numerically.
%To  do this we draw 10,000 random points from a Gaussian with $\langle x\rangle=\langle y\rangle=0$, and a variance $\sigma_{\rm w}$.
%We then shift these randomly with x and y widths according to the empirical error in Table \ref{tab:errors} and then convert %these into proper, radial distances. 
%We repeat this 10,000 times for each redshift in the survey sample generating a total of 100,000 points. 

Using these posteriors we can compare the distribution of the 10 maximum likelihood positions returned from the MCMC in {\sc lenstool}, and compare this distribution to each posterior.
We then have for each posterior a probability that the distribution of positions were drawn from it, and hence generate best fitting value of $A_{\rm w}$. The error in this is just when the likelihood that the measured distribution is drawn from a particular posterior falls below the $1\sigma$ threshold.

%\fig
%\includegraphics[width=0.5\textwidth]{bahams_wobbling.pdf}
%\caption{ We measure the BCG wobble in the BAHAMAS simulations. We analyse both the low resolution sample consisting on 600 clusters, and the high resolution sample of 22 clusters. The black histogram and associated dashed line represent the offsets in the low res sample and its best fit to the numerical posteriors. The grey histogram and associated dashed line show the distribution of offsets in the high res simulation and its best fit posterior. The legend shoes the wobble amplitude for these best fit lines and their associated errors. The stark difference values show lack of convergence in the high resolution sample, despite the low number. 
%\label{fig:bahams_wobble}}
%\efig	

\section{Results}\label{sec:results}

\fig
\includegraphics[width=0.5\textwidth]{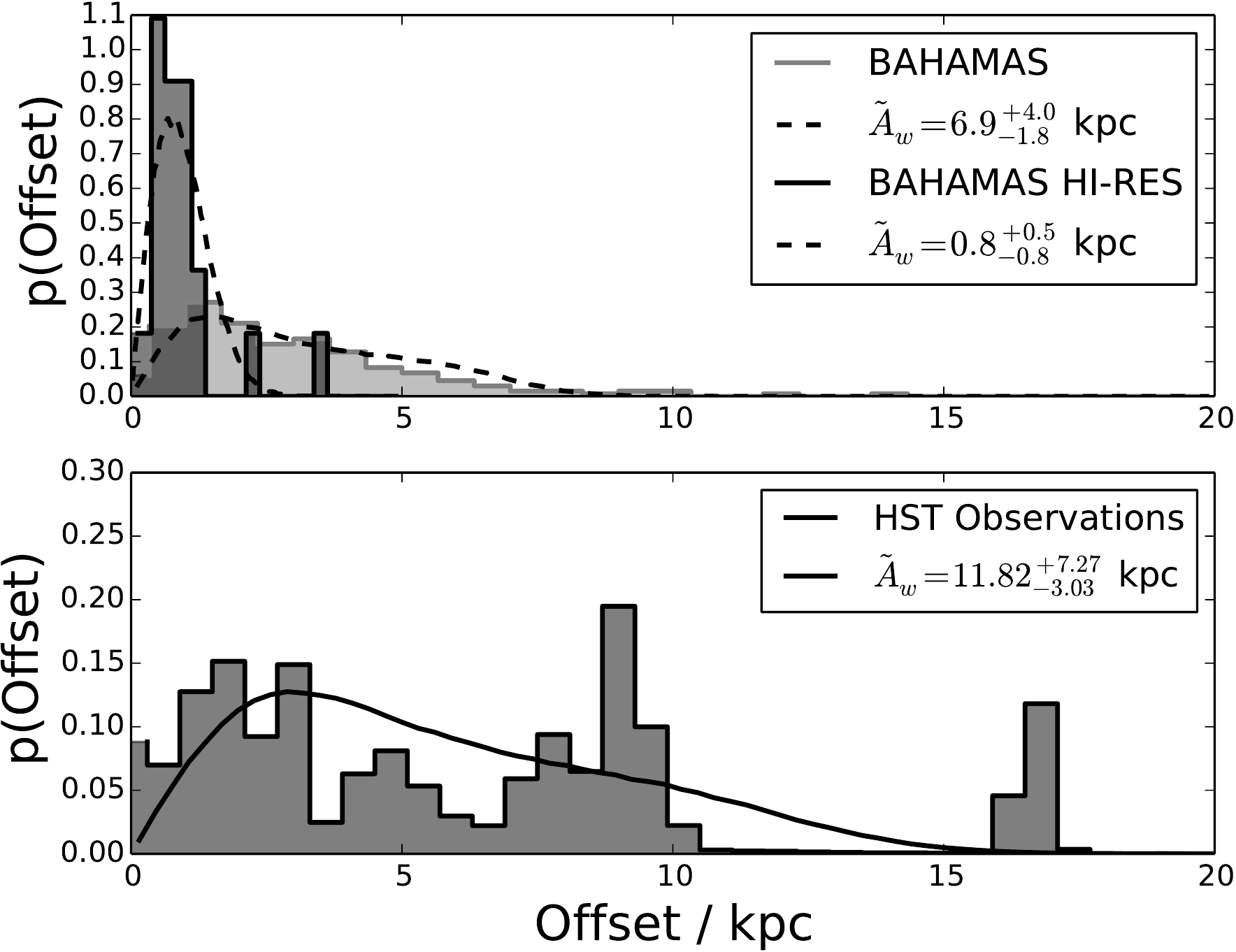}
\caption{ The top panel shows the distribution of radial offsets between the BCG and the dark matter potential in the BAHAMAS simulations. We analyse both the fiducial resolution sample consisting on 600 clusters, and the high resolution sample of 22 clusters. The dark histogram and associated dashed line represent the offsets in the high-resolution sample and its best fit to the numerical posteriors. The light-grey histogram and associated dashed line show the distribution of offsets in the high-resolution simulation and its best fit posterior. The legend shows the wobble amplitude for these best fit lines and their associated errors. The bottom panel shows the distribution of samples in the MCMC from lenstool from the ten mass reconstructions of the observed galaxy clusters. The fitted line is the best fit posterior to the distribution of {\it best fit positions}. The legend shows the wobble amplitude of this best fit line with its associated error. 
\label{fig:observation_wobble}}
\efig	

\subsection{Image simulation results}
In section \ref{sec:errors} we generated 5 sets of image simulations to empirically measure the error in {\sc lenstool}.
In each suite we introduced a fake wobble amplitude of $A_{\rm w} = [0, 1, 5, 10, 20]$~kpc.
Here we present the results from measurement of the wobble amplitude 
%For each input wobble amplitude we analyse the image simulation of ten galaxy clusters to check the robustness of the method

%We first take the simulations used in section \ref{sec:errors} where we simulate the ten observed clusters, however induced a known oscillation amplitude, to see if we can recover the input oscillation amplitudes and test the robustness of our method.

%We take all the points in the posterior sampled during the MCMC in the x and y space, and create a radial posterior centred on the main halo. 
%We then stack this radial posterior for the complete sample of galaxy clusters to get a posterior of the total offset between the main halo and the BCG.
%We carry out a KS-test on the distribution of best fit position values (and not the whole posterior).
Figure \ref{fig:compare_posteriors} shows the posteriors derived from the observation simulations. 
In each panel we show a different image simulation stacked posterior in the coloured histogram, and the best-fit numerically calculated posterior from Figure \ref{fig:posteriors} according to the KS two-sample test as the solid black line.
The key shows the input wobble amplitude, $A_{\rm w}$, and the estimated wobble amplitude, ${\tilde A}_{\rm w}$ that corresponds to the solid black line with its associated uncertainty.
We generate the uncertainty by finding the wobble amplitude at which probability $P$ given the data, $D$ is less than 0.32 (i.e. $P(A_{\rm w}|D)<0.34$).
We find that in each simulation, the estimated wobble amplitude is consistent with the input at the $1\sigma$ level. 

\subsection{Cosmological Simulation Results}
We next estimate the BCG wobble amplitude in the BAHAMAS simulations. 
We measure the radial distribution of the offsets between the BCG and the centre of the cluster potential using fiducial and high-resolution simulations.
We then estimate the wobble amplitude by comparing these distributions to a variety of posteriors calculated numerically for a given $A_{\rm w}$. The top panel of Figure \ref{fig:observation_wobble} presents the two distributions plus their best fitting numerically calculated posteriors, with their associated wobble amplitude in the legend. We find that the fiducial resolution simulations exhibit a radial distribution consistent with a wobble amplitude of $A_{\rm w}=6.9^{+4.0}_{-1.8}$kpc. 
However, we note here that the model with which we fit to the data to derive the offsets is parameterised by a wobble amplitude and the expected random error in the model fitting process. This  does not allow for an extra degree of error associated with the smoothing length of the simulations since the real data does not have such a thing. As mentioned, the smoothing length of the fiducial simulation is 4$h^{-1}kpc$ which is almost exactly what is found with out model. We can include this into the error model such that the error stated includes this, or we use the exact same model on the simulations as we do in the data, we decide to use the latter.

We therefore carry out the same experiment on the high-resolution simulation. We find that the high-resolution simulations exhibit a radial distribution consistent with a much smaller amplitude wobble of $A_{\rm w}=0.8^{+0.5}_{-0.8}$kpc.  We therefore find that the high-resolution simulations predict an offset distribution that is consistent with zero\footnote{The trend that the offset distribution becomes narrower and closer to zero as the resolution of the simulation is increased appears to be consistent with the findings of \citet{LCDM_offsets}, who used the EAGLE simulations, which are approximately a factor of 4 higher in spatial resolution than BAHAMAS hi-res, to explore the offsets between stars and dark matter.  \cite{LCDM_offsets} find an offset of $\la 200$ pc.  A caveat in making this comparison, however, is that the EAGLE sample does not include massive clusters, given the relatively small volume, and \citet{LCDM_offsets} did not compute galaxy centres in an observational manner (i.e., from analysis of imaging data).} and has a upper limit of $A_{\rm w}<2$kpc at the 95\% confidence limit.  

\subsection{HST observation results}
Finally we carry out the same test on the observations. We take the best fit positional parameters and construct a radial offset distribution. We then find the posterior that describes this distribution the best.
In Figure \ref{fig:observation_wobble} we present the stacked posterior from all the reconstructions and the best fit posterior with its associated wobble amplitude, $A_{\rm w}$ shown in the legend.
We find that the data has a amplitude of $A_{\rm w}=11.8^{+7.3}_{-3.0}$kpc and is ruled out as a zero wobble at the $3\sigma$ level (p value = 0.005).
%Table \ref{tab:kstest} presents a full overview of the results from all the KS tests. 
%\begin{table}
%\centering
%\begin{tabular}{|c|c|c|c|}
%\hline
%Dataset & $\tilde{A}_{w}$ & p($\tilde{A}_{\rm w})$ & p$(A_{\rm w}=0)$ \\ 
%\hline\hline
%$A_{\rm w}=0$kpc&    $0.2^{+5.9}_{-0.2}$ & 0.99 &0.985 \\ 
%$A_{\rm w}=1$kpc &   $0.2^{+3.5}_{-0.2}$ & 0.90 &0.900 \\ 
%$A_{\rm w}=5$kpc &   $3.5^{+5.0}_{-2.1}$ & 0.75  &0.280 \\ 
%$A_{\rm w}=10$kpc & $8.8^{+4.8}_{-3.0}$ & 0.99 &0.014 \\ 
%$A_{\rm w}=20$kpc&  $16.6^{+9.2}_{-4.6}$ & 0.98 & 0.001 \\ \hline
%BAHAMAS & $6.6_{-1.5}^{+1.5}$ & 0.38 & 0.0\\ 
%BAHAMAS HI-RES & $0.8_{-0.8}^{+0.5}$ & 0.90 & 0.6\\ \hline
%HST Observations &                         $11.82^{+7.3}_{-3.0}$ & 0.96 & 0.005 \\ \hline
%
%\end{tabular}
%\caption{\label{tab:kstest} The results from comparing various datasets to those posteriors in Figure \ref{fig:compare_posteriors} using the  KS two-sample test. {\it Col 1:} the name of the data / simulation, {\it Col 2:} the wobble amplitude with the highest probability and its $1\sigma$ upper and lower uncertainty, {\it Col 3:} the probability of the best fit wobble amplitude, {\it Col 4:} the probability of wobble amplitude, $A_{\rm w}=0$kpc given the dataset.}
%\end{table}
\section{Sensitivity to model choice}
Throughout our study we have adopted an NFW as our fiducial density profile for the main cluster halo.
This was valid under our null hypothesis.
However evidence for a core does raise the question of model dependency. We therefore choose to test our results with a pseudo isothermal elliptical mass distribution (PIEMD). This mass profile has the additional degree of freedom in the core, better reflecting a profile in a SIDM scenario. 
We carry out the same test as previously, measuring the positions of each halo with respect to the BCG. Figure \ref{fig:piemd} shows how the distribution and best fit posterior is consistent with the NFW. We also find no significant difference in the quality of fit between the two profiles. This is encouraging, not only showing that the results are not sensitive to the choice of profile, but also that this method of measuring a core is more sensitive than simply measuring it. 
\fig
\includegraphics[width=0.5\textwidth]{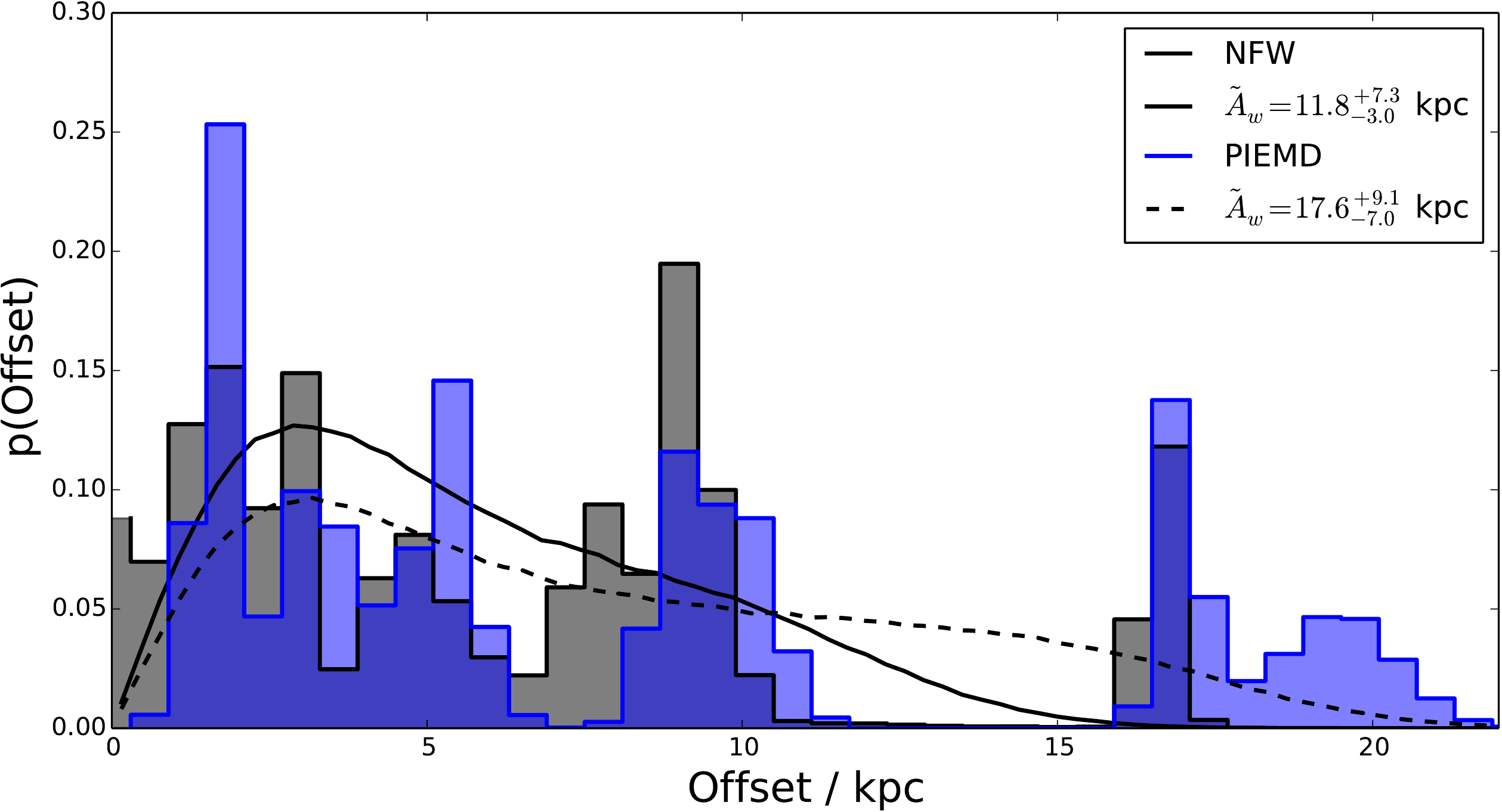}
\caption{ Distribution of offsets for the fiducial dark matter model, NFW, in grey, and its best fitting posterior in the solid black line and the corresponding wobble in the legend. In blue are the offsets when fitting a PIEMD dark matter model which has an extra degree of freedom in the core, with the best fitting posterior and its associated wobble in the legend.
\label{fig:piemd}}
\efig	

\section{Conclusions}\label{sec:conclusions}
Extensions to the cold dark matter model predict the existence of cores in the centres of galaxy clusters, something not apparent in standard model of dark matter.
Simulations of collisional dark matter predicts that the Brightest Cluster Galaxy (BCG) in a relaxed halo that has experienced a major merger in its formation will exhibit a residual `wobble' due to the existence of a core \citep{darkgiants}, whereas no such wobble is found in standard model dark matter.

In this paper we test this hypothesis using the BAHAMAS suite of cosmological simulations \citep{BAHAMAS} and observations of ten relaxed galaxy clusters from the CLASH and LoCuSS surveys.
Using the public software {\sc lenstool}, we model the strongly lensed images of distant galaxies to measure the radial offset between the large scale cluster halo and the BCG and ask whether we observe any evidence for an excessive variance that cannot be accounted for purely with positional uncertainties including a realistic account for systematics. 

We estimate the wobble amplitude in the BAHAMAS fiducial and high-resolution simulations. 
We find that the result is sensitive to the adopted resolution, in the sense that the offset distribution becomes more narrow and closer to zero as the resolution is increased.  The BAHAMAS high-resolution simulations indicate $A_{\rm w}=0.8_{-0.8}^{+0.5}$. We therefore conclude by placing an upper limit wobble amplitude from the simulations of $A_{\rm w}<2$kpc at the $2\sigma$ confidence limit.  This likely represents a conservative upper limit, given that the results are not converged to resolution (see also \citealt{LCDM_offsets}).

We empirically estimate the errors on the strong lensing model, by creating a suite of ten image simulations. 
For each image simulation we use a known analytical model of a galaxy cluster to produce multiple images that closely reflect the distribution of multiple images in the observed, real data.
In each case the cluster halo is centred exactly on the BCG.

We find that when the cluster halo is exactly coincident with the BCG, with no offset, we observe an uncertainty in the radial position of the offset of $\sigma_{\rm r}\simeq0.7\arcsec$ or $\sigma_{\rm r}\simeq3.1$~kpc (taking into account the distribution of cluster redshifts used in this study). If we compare this to the mean uncertainty reported by {\sc Lenstool} of $\sigma_r\simeq0.05\arcsec$ ($0.22$~kpc), we find that the modelling procedure significantly underestimates the uncertainty in the position of the halo.
Following this test we create 4 more suites of image simulations, each with a ten galaxy clusters, now containing fake input signal.
We model the BCG wobble as a simple harmonic oscillator, described uniquely by its wobble amplitude, $A_{\rm w}$.
We simulate four wobble amplitudes up to 20~kpc and find that our method recovers each case within the 68\% confidence region.

%To derive estimates of an unkownn wobble amplitude, we numerically calculate multiple posteriors, each for a given wobble amplitude $A_{\rm w}$, and error in the position of  $\sigma_{\rm r}\simeq0.7\arcsec$. Using the Kolmogorov-Smirnov two-sample test (KS test), we then find for a given radial offset distribution the most likely oscillation amplitude.
%We apply this method first to the suite of four simulations with input, fake wobble amplitudes. In all cases we recover the input value within the 68\% confidence region.

Finally we estimate the wobble amplitude within ten relaxed galaxy clusters observed by the Hubble Space Telescope. We find that the data prefers a wobble amplitude of $A_{\rm w}=11.82^{+7.3}_{-3.0}$~kpc, and disfavours a zero-wobble amplitude at the $3\sigma$ confidence level.

One effect that could cause some variance in the position of the centres of the clusters is line of sight structure. 
\cite{losoffsets} found that weak lensing positional peaks can be offset by $\sim$5kpc, where the mass of structure along the line of sight would be comparable to the size of the halo being studied. In this sample we measure the positions of ten clusters with a mean mass of over $10^{15}$M$_\odot$, and hence would require a very large halo along the line of sight to perturb the position of the cluster. The chances of having two clusters of $M\sim10^{15}M_\odot$ along the same line of sight is very small \citep{Harvey13}, and hence we argue that the variance we see is not due to line of sight structures.

Simulations by \cite{darkgiants} predicted that the amplitude of a BCG wobble would correspond directly to the size of a constant density core in the centre of a galaxy cluster and ``decayed minimally". This study therefore finds evidence for a mean finite core of $\sim 11$kpc at the centre of galaxy clusters,  potentially indirectly inferring the need for new physics. 
Interestingly, this is consistent the direct measurement of cores by  \cite{densityProf2,densityProf3}, where they found a mean core size of $\langle\log r_{\rm core}\rangle = 1.14\pm0.13$~kpc. Moreover, it also shows the importance of not assuming that the BCG lies at the centre of a galaxy cluster when fitting parametric fits in strong lensing reconstructions. \cite{darkgiants} claim that is would equate to a cross-section of $\sigma_{\rm DM}/m \sim 0.1$cm$^2$/g.
However, this will need to be confirmed with more data as increased observations of relaxed clusters are observed, and the consequences need to further modelled using simulations.  Surveys such as Euclid \citep{Euclid}, which will provide thousands of relaxed clusters with strong lensing information will allow us to not only confirm this finding, but also probe the redshift evolution of the core.

\section*{Acknowledgments} This research is supported by the Swiss National Science Foundation (SNSF). D. Harvey also acknowledges support by the Merac foundation.  The authors thank Joop Schaye for his contributions to the BAHAMAS simulations, Richard Massey and Mathilde Jauzac for extremely useful conversations and finally Johan Richard and Marceau Limousin for providing multiple image positions and {\sc Lenstool} parameter files for the cluster reconstructions.

\bibliographystyle{mn2e}
\bibliography{bibliography}

\bsp

\label{lastpage}

\end{document}